%% file: main.tex
\documentclass[conference]{IEEEtran}

\usepackage{todonotes, tabularx, graphicx, multirow, booktabs, subcaption}
\usepackage{xparse} 
\NewDocumentCommand{\rot}{O{45} O{1em} m}{\makebox[#2][l]{\rotatebox{#1}{#3}}}%
\usepackage[utf8]{inputenc}

\title{Bursts of Activity: Temporal Patterns of Help-Seeking and Support in Online Mental Health Forums}

\author{\IEEEauthorblockN{Taisa Kushner}
\IEEEauthorblockA{
\textit{University of Colorado Boulder}\\
Boulder, USA \\
taisa.kushner@colorado.edu}
\and
\IEEEauthorblockN{Amit Sharma}
\IEEEauthorblockA{
\textit{Microsoft Research}\\
Bangalore, India \\
amshar@microsoft.com}
}

\newcommand\copyrighttext{%
  \footnotesize This is an author pre-print. Final publication is available in the Proceeding of the Web Conference 2020 (WWW '20)  https://doi.org/10.1145/3366423.3380056}
\newcommand\copyrightnotice{%
\begin{tikzpicture}[remember picture,overlay]
\node[anchor=south,yshift=10pt] at (current page.south) {\fbox{\parbox{\dimexpr\textwidth-\fboxsep-\fboxrule\relax}{\copyrighttext}}};
\end{tikzpicture}%
}

\begin{document}
\maketitle
\copyrightnotice

\begin{abstract}
    Recent years have seen a rise in social media platforms that provide peer-to-peer support to individuals suffering from mental distress. 
    Studies on the impact of these platforms have  focused on either short-term scales of single-post threads, or long-term changes over arbitrary period of time (months or years). While important, such arbitrary periods do not necessarily follow users' progressions through acute periods of distress. Using data from Talklife, a mental health platform, we find that user activity follows a distinct pattern of high activity periods with interleaving periods of no activity, and   
 propose a method for identifying such \emph{bursts} \& \emph{breaks} in activity.   
We then show how studying activity during bursts can provide a personalized, \emph{medium-term} analysis for a key question in online mental health communities: What characteristics of user activity lead some users to find support and help, while others fall short? Using two independent outcome metrics, moments of cognitive change and self-reported changes in mood during a burst of activity, we identify two actionable features that can improve outcomes for users: persistence within bursts, and giving complex emotional support to others. Our results demonstrate the value of considering bursts as a natural unit of analysis for psychosocial change in online mental health communities.

\end{abstract}

\section{Introduction}
\input{intro}

\section{Related Work}
\input{relatedwork.tex}

\section{Data}
\input{data}\label{sec:data}
\section{Bursts and Breaks in User Behavior}\label{sec:identify-burst}
\input{identifying_bursts.tex}
\section{Burst Features and MOC}\label{sec:features}
\input{features.tex}

\section{Burst Features and Change in Mood}\label{sec:moodchange}
\input{user_benefit.tex}
\section{Concluding Discussion}
\input{implications_discussion.tex}

\bibliographystyle{plain}
\bibliography{references}

\end{document}

%% file: intro.tex
Common mental health disorders like depression and anxiety affect millions of people across the world. The World Health Organization estimates that by 2030 depression will be leading cause of the global disease burden~\cite{lancetdepression}, surpassing  cardiovascular disease. However,  in both the Global North and Global South, access to mental health is a concern~\cite{patel2018lancet}. In many parts of the world, there is also a stigma associated with mental illness. 

As a result, online mental health platforms are becoming popular places where people support each other in times of mental distress~\cite{naslund2016future}, often using a pseudonym-based identity. For example, platforms such as Talklife and 7Cups have thousands of users globally who come on to the platform to seek and provide mental health support. A  key distinction of these platforms from those which offer online therapy is that they depend largely on peer supporters who are often not trained in counselling, unlike other mental health services.  This has prompted recent research into various facets of peer-to-peer support and how it may help or hinder a person in mental distress. These works have taken two main directions to understand peer support: short-term, looking at support behavior on a single conversation thread~\cite{Pruksachatkun2019} and  long-term, looking at longitudinal changes in psychological well-being for individuals over years~\cite{de2016discovering,park2017longitudinal}.  

In this paper, we extend this line of work and study the medium-term changes in well-being that can provide a unique perspective on \emph{how} mental health changes happen over time. 
Specifically, we argue for the need to consider medium-length interactions which occur as a series of posts, replies, and threads clustering around a specific topic or user's state of increased engagement with the platform, rather than considering only changes which occur within individual threads or over arbitrary lengths of time such as weeks, months or years.  

Using data from an online mental health forum, Talklife, we find that user's medium-term interactions follow a striking pattern: \emph{bursts} of high activity with interleaving \emph{breaks} of no activity,  unlike usage of typical online social platforms~\cite{guo2009userpatterns}.  These bursts can be considered as proxies for individuals' progress through an acute time of distress or their desire to help others by increasing engagement with the platform.  
Thus, we posit bursts as a natural unit of analysis for well-being on an online mental health forum. Compared fixed-short term analysis such as sessions~\cite{jansen2007definingsession} of web search and other websites~\cite{halfaker2015usersessions}, and long-term analyses of changes in behavior, studying medium-term bursts  can  provide a fine-grained understanding on how and what kinds of peer support and user's actions on the platform lead to positive changes in an individual's psychosocial well-being.


While intuitive, defining a burst is non-trivial due to variability in different individuals' time-scales of activity.  We propose the definition of \emph{burst} as a personalized metric related to an individual's median inter-post time. We find that a majority of the users on Talklife engage
with the platform in bursts and breaks. Bursts are substantially different from  sessions~\cite{halfaker2015usersessions} commonly used to study user activity: the metric is personalized, and an average burst consists of $20$ posts across $6.5$ days.

To demonstrate the value of measuring activity via bursts, we then study how the support experience of individuals is shaped by interactions within their bursts and breaks on the Talklife platform. 
 In particular, we compare bursts
that were successful in leading to a positive mental health change
with those that were not, in order to identify features of activity
that are associated with positive change. For robustness, we utilize two measures of well-being: moments of positive cognitive change~\cite{Pruksachatkun2019} where people self-report a
positive cognitive change, \& changes in their self-reported mood when posting. We find that users who undergo a positive cognitive change
over a burst are more likely to engage with others at a
higher rate through posting replies on other’s posts, participate in increased complex support and lower simple support when replying to others, and have increased post diversity while maintaining similarity between the categories they post replies and original posts in.


Our results have relevance for online communities, in particular with how we think about user interactions with online mental health platforms and user churn and retention on such platforms. 
To summarize, our contribution include: defining ``bursts'' and ``burstiness'' of user activity as it relates to medium-length interaction on a mental health platform, identifying features which correlate with users experiencing a cognitive moment of change within a burst, and robustly testing these features against self-reported changes in mood to determine two actionable suggestions for improving user experience: persistence within a burst, and giving complex emotional support to others.

%% file: relatedwork.tex
Our work draws on temporal analysis of activity in online communities and understanding of psycho-social well-being in mental health forums.  


\subsection{Temporal analysis in online communities}
Temporal patterns of user activity are useful for understanding various aspects of online community dynamics, such as growth of a community~\cite{kumar2010structure}, diffusion of information~\cite{guille2012predictive},  change in popularity of items~\cite{yang2011temporalpatterns} or expertise of users~\cite{pal2012evolutionexperts}. They have also been used to model users' preferences for items~\cite{jamali2011modeling}. When considering frequency of posting, in typical online social networks, studies have found that users follow a distinct daily and weekly pattern~\cite{guo2009userpatterns,golder2007rhythms}. Long-term modeling of activity often uses a uniform discrete time-step process~\cite{kumar2010structure,jamali2011modeling}, thereby ignoring the variability in time between activities.

Perhaps the closest to our work is the literature on studying increases (bursts) of activity in social media due to external events. Tamime et al.~\cite{al2018wikiburst} look at the burstiness in editing activity on Wikipedia due to disease outbreaks, and define burstiness by comparing it to concurrent activity on other articles. On Twitter, Zhao et al.~\cite{zhao2012eventbursts} describe a method to detect activity bursts due to an external event.  Kumar et al.~\cite{kumar2005bursty} found a bursty pattern in the edges created for different blog communities. However, the common notion of burst in these studies is a spike in activity due to external events whereas our goal is to identify periods of sustained usage (bursts) or non-usage, that are shaped by people's own mental health state. 

In this sense, our findings most closely mimic health-related searches from the activity logs of a web search engine~\cite{white2009cyberchondria}. Similar to White and Horovitz's work~\cite{white2009cyberchondria}, we extend the literature on bursts by considering burst as a period of sustained usage. In Section~\ref{sec:identify-burst}, we discuss how our proposed method for identifying bursts differs from past methods motivated by text or event streams~\cite{kleinberg2003bursty,fung2005parameter}. 

\paragraph*{Churn and Retention} Results from this work also extend to churn \& retention.
While extensive work has been done to study user churn in online communities~\cite{pudipeddi2014user,dror2012churnyahoo}, the focus has been mainly on temporal patterns of usage, such as recency \& frequency~\cite{wei2010reviewrfm}, or social network-based features~\cite{dasgupta2008social}. However, here too, models for churn do not take into account the bursty nature of activity, if present. Depending on the churn cutoff, a long inter-burst time may be interpreted incorrectly as a churn event. Considering bursty behavior of users allows us to develop a more accurate understanding of churn in an online community.

\subsection{Changes in Well-Being in Mental Health  Forums} The increasing popularity of online mental health forums has prompted work on expression and support interactions around mental health. While initial work focused on prediction of mental health state through individuals' posts~\cite{de2013predicting,tsugawa2015recognizing}, recent work focuses on the quality of support received on these forums. Studies have looked at how factors such as extent of self-disclosure~\cite{ernala2017linguistic}, anonymity~\cite{de2014mentaldisclosure}, linguistic accommodation~\cite{sharma2018linguistic}, and differences in culture~\cite{Pruksachatkun2019}  affect supportive interactions on forums such as Reddit and Talklife. Andalibi et al.~\cite{Andalibi2017} provide a taxonomy of different kinds of support, such as emotional support, that we use for classifying support interactions within bursts.  

For outcomes on online mental health communities, previous work has looked at mental health outcomes either the level of a single thread, or longitudinal analysis on time scales of months or years. Pruksachatkun et al.~\cite{Pruksachatkun2019} analyzed support comments on a thread and built a model for predicting whether the thread leads to a positive cognitive change for the individual. To measure a positive cognitive change, they used a dictionary of phrases where people self-report their change.  
On larger time-scales, longitudinal studies look at long-term changes in psychosocial outcomes due to the nature of peer support received on an online mental health forum.  De Choudhury et al.~\cite{de2016discovering,de2017language} analyze indicators that lead people towards suicidal ideation and identify terms used in support responses on a mental forum in Reddit that lead to the highest causal effect on a positive outcome. In another study on Reddit, Park and Conway~\cite{park2017longitudinal} find that linguistic indicators of psychosocial outcomes such as positive emotion increase with prolonged use of the community.  

However, little is known how these changes happen at the medium term. Our work looks at support behavior and its associated effects at the medium time-scale using bursts as a unit of analysis.

%% file: data.tex
Longitudinal data of user activity and interactions was acquired from the online peer-to-peer support network, Talklife. \cite{talklife} 
The platform is set up as a chronological feed with original posts \& subsequent comments (replies) on posts either from others or the original user. An \emph{original post} along with the \emph{replies} on that post is collectively  referred to as a \emph{thread}.

Users can choose to post original posts using their (self-selected) user name or anonymously. Original posts are associated with two main features:  \textbf{category} and \textbf{mood}.

\begin{figure*}[t]
    \centering
    \includegraphics[width=0.48\textwidth]{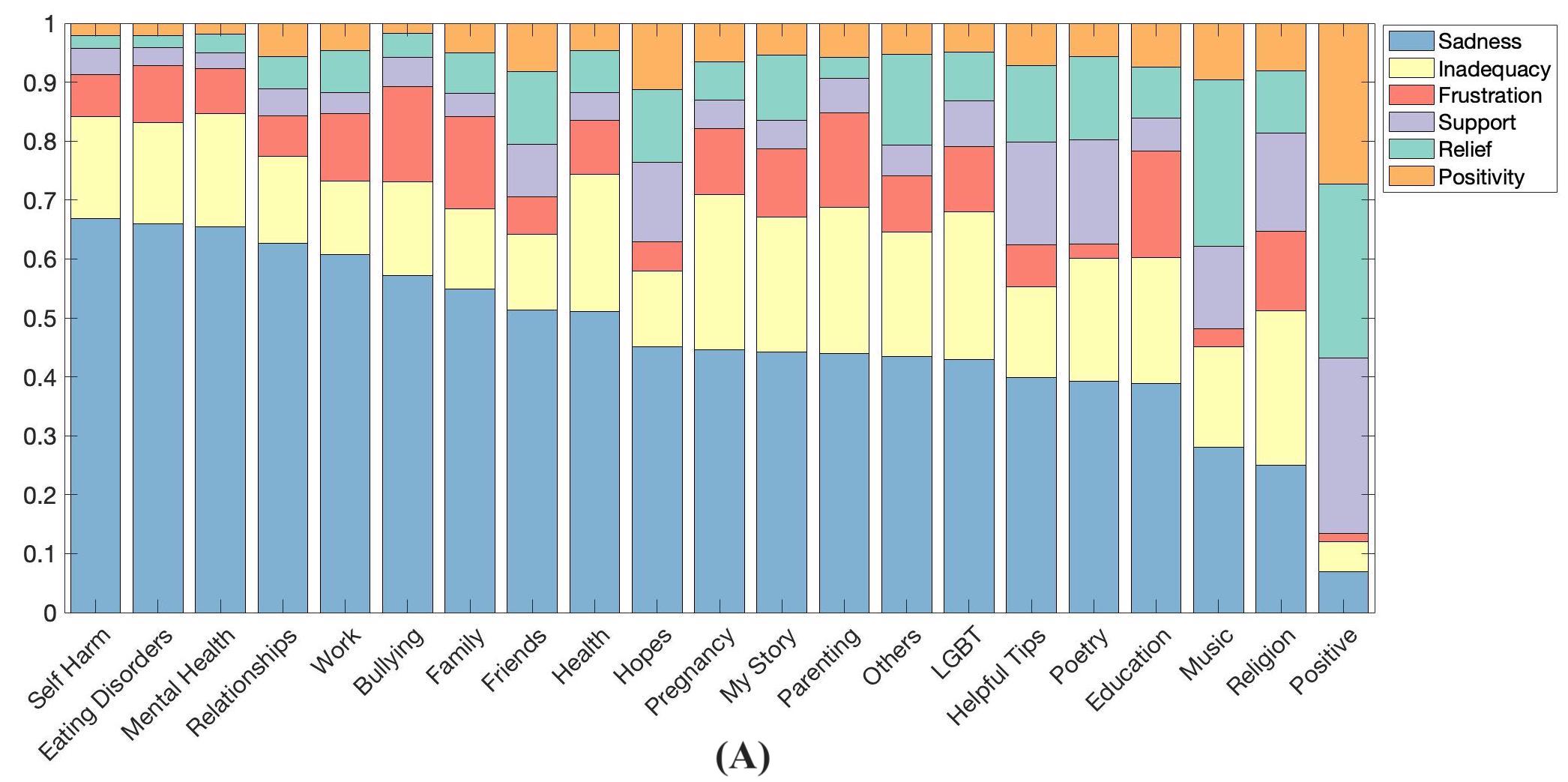}
    \includegraphics[width=0.48\textwidth]{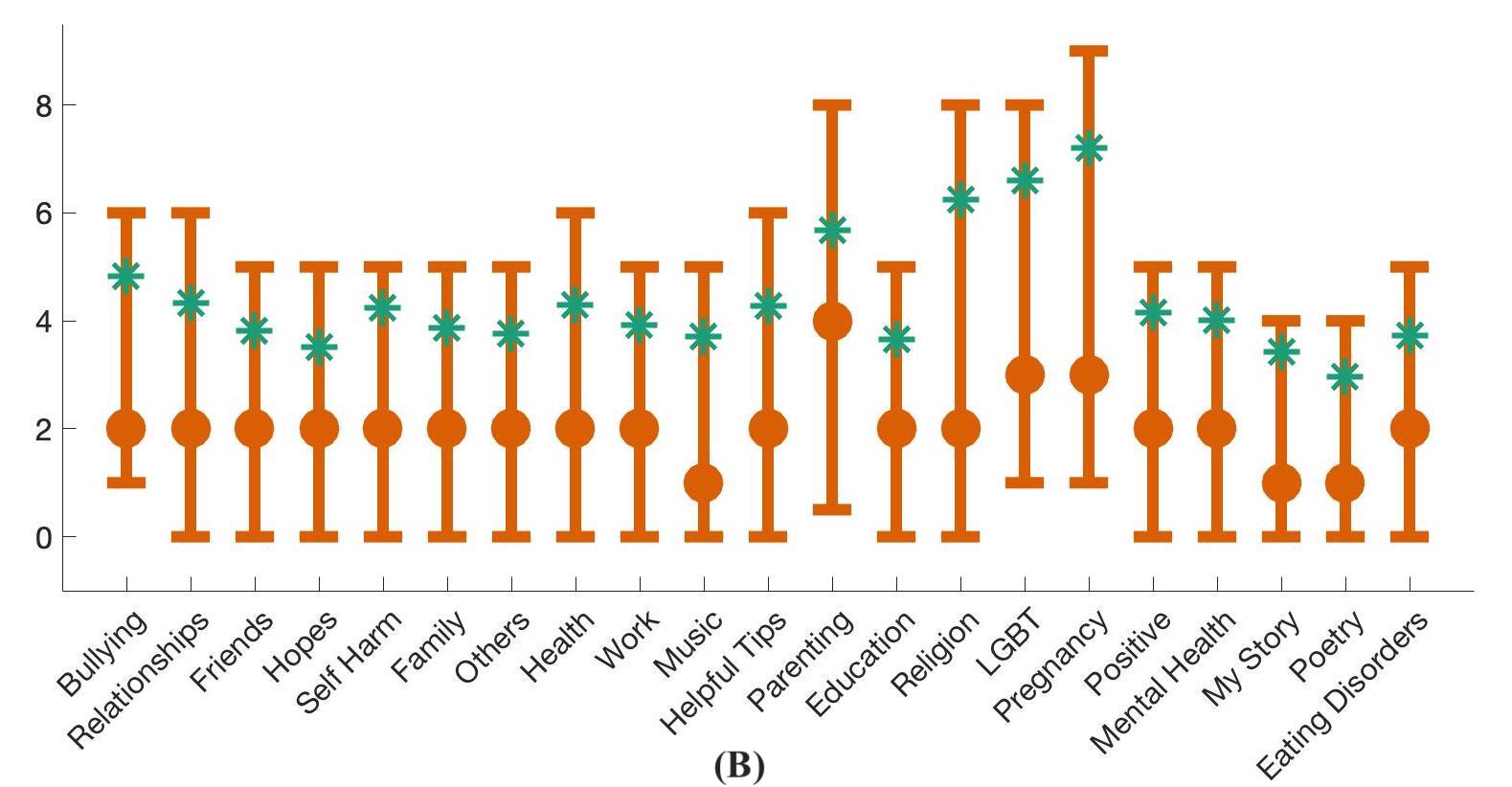}
    \caption{Posting activity by Category. (A) Mood distribution of posts within each category (B) Median (orange circle), mean (green star), 25$^{th}$ and 75$^{th}$ quantile ranges for number of replies per post, per category.}
    \label{fig:stats_by_cat}
\end{figure*}

 \paragraph*{Categories} These are user-selected and range in specificity from the narrower category "Parenting" to the all encompassing "Others"  (Fig.~\ref{fig:stats_by_cat}). Nearly $20\%$ of posts fall into the Relationships category, $17\%$ into the category Others, and $8.5\%$ into each of Mental Health, My Story, Self Harm, and Friends. Least posted categories include Eating Disorders, Pregnancy, Religion \& Parenting, with $<1\%$ each. 

 \paragraph*{Mood} Users must select one of 59 "moods" when posting. These moods are grouped by Talklife into six thematically related pages consisting of words which evoke similar feelings. For ease of interpretation we give these groups descriptive names (Table~\ref{table:moodgroups}). Note these mood groups encompass a wide range of moods.
\vskip 0.04in
\begin{table}[t]
    \centering
    \footnotesize
        \caption{Mood groups, associated moods \& percentages of overall posts. Most-frequently used word/group is bolded.}
\begin{tabular}{|p{1.4cm}|p{5.5cm}|p{0.35cm}|}
    \hline
      \textbf{Group}  & \textbf{Included Moods} & \textbf{Pct}\\
     \hline
    \textbf{Sadness} & \textbf{sad}, heartbroken, depressed, anxious, nervous, down, lonely, tired, insecure, exhausted, overwhelmed, afraid & 52\% \\
    \hline
    \textbf{Inadequacy} & worried, \textbf{meh}, inadequate, numb, confused, embarrassed, shocked,sick, bored, nothing & 17\% \\
    \hline
    \textbf{Frustration} & \textbf{frustrated}, annoyed, angry, furious, irritated, jealous, stressed, moody, disgust & 8\% \\
    \hline
    \textbf{Support} & \textbf{supportive}, hopeful, optimistic, loving, inspired, proud, nostalgic, caring, loved, supported & 7\% \\
    \hline
    \textbf{Relief} & excited, amused, thankful, \textbf{calm}, relaxed, chilled, relieved, jolly, determined, motivated & 10\% \\
    \hline
     \textbf{Positivity} & astonished, \textbf{positive}, surprised, encouraged, happy, amazed, ecstatic, energetic & 6\% \\
    \hline
\end{tabular}
\label{table:moodgroups}
\end{table}
Our goal is to monitor user's progressions through these groups, from ``negative'' categories 1-3 to ``positive'' categories 4-6.
Fig.~\ref{fig:stats_by_cat}A shows the association between moods and categories of original posts. Categories such as Self-Harm and Eating Disorders have the highest proportion  of ``sadness'' mood posts while categories such as Music, Religion and Positive have the lowest proportion. Across all categories except Positive, sadness predominates. 

 \paragraph*{Replies}
 Reply posts occur when a user replies to (comments on) an original post, and cannot occur anonymously. Any replies, be it by other users or by the user who posted the original post, are referred to as "replies" and are tied to the original post in what we refer to as a "thread". Reply posts have no affiliated "mood" labels, \& are tagged in the same category as the original post within the thread. On average, posts receive $4.39\pm 1.13$ replies. Notable is that some of the least posted-in categories of Parenting, Religion, Pregnancy and LGBT, have highest overall reply numbers (Fig.~\ref{fig:stats_by_cat}B).

 


\paragraph*{Filtering}
For our study, we analyze longitudinal  data from a user's very first post to the end of our data set, Jan. 14, 2019.  For a given user, their timeline will include in chronological order: (1) all original posts, (2) reply threads for each of their original posts, if present, and (3) replies the user left on other's posts.
We consider full timelines from a randomly selected sub-sample of 3000 users on the Talklife platform who meet the following criteria: 
\begin{itemize}
    \item A "regular" user, not an administrator/platform moderator
    \item Joined after the most recent update of "mood" categories, Dec. 15 2015
    \item Minimum of 1 original post and 1 reply posted
    \item Minimum of 10 total posts (original posts or replies)
    \item Maximum of 2028 posts (original posts or replies)
\end{itemize}
The final condition on maximum posts is based on the three-sigma rule of thumb, a the standard heuristic for encompassing 99\% of all individuals while removing any extreme outliers \cite{grafarend2012}. 
In total, our data set includes nearly 500,000 posts by 3000 users with an average \emph{active} account age of $168$ days. The active account age for each user is computed as (most recent post) - (first post).

%% file: identifying_bursts.tex
In terms of community structure, Talklife lies between an unstructured timeline-based platform (e.g., Facebook) \& a question-based online community (e.g., mental-health subreddits on Reddit). 
While the platform is centered around mental health, users view others' posts as a chronological timeline rather than questions or topics with titles.
Thus, posts vary from structured, detailed questions, to ``status updates'' such as (paraphrased) "\textit{The leaves are crunchy today}". 
As noted in Section.\ref{sec:data}, while some specific categories exist (e.g., Parenting), almost $20\%$ of posts end up in the "Others" category. This suggests that users either often cannot find a relevant category for their question or do not place significant weight on categories when interacting with the platform. 

The above factors lead to a unique platform where users post a series of \emph{original} posts as questions or follow-ups centered around a single acute problem. For example, a user may begin with a post about troubles with a partner, asking the community for advice. Then the user may continue to post on this topic for a couple of days, either receiving advice \& support from community members, or not. Depending on past interactions, a user will often follow-up positively: (paraphrased) "\textit{Thanks you guys, your advice on my last post was very helpful. I was able to talk to my partner and it wasn't as bad as I thought it would be.}, 
or negatively: (paraphrased)
"\textit{Talklife isn't helpful.}".
After this, a user will tend to take a break from the platform---possibly due to a resolution to their topic or "giving up"---and return to the platform after a while with another, often also acute problem.
Upon observing this pattern for numerous users, we sought to systematically identify these patterns and quantify the extent to which users' behavior follows this pattern.

\subsection{Formal Definition}
We formalize the concepts of activity \emph{bursts} \& \emph{breaks} based on an individual's median inter-post time, which is defined as the median time taken between all posts (original or replies). 
\vskip 0.04in
\noindent \textbf{Activity Burst:} A period of sustained activity on the platform by posting original posts or replies such that the time between any two consecutive posts remains within $N*(median$ $user$ $inter$-$post$ $time)$, where $N$ is a sufficiently large number.
\vskip 0.04in

\begin{figure}[t]
    \centering
    \includegraphics[width=0.45\textwidth]{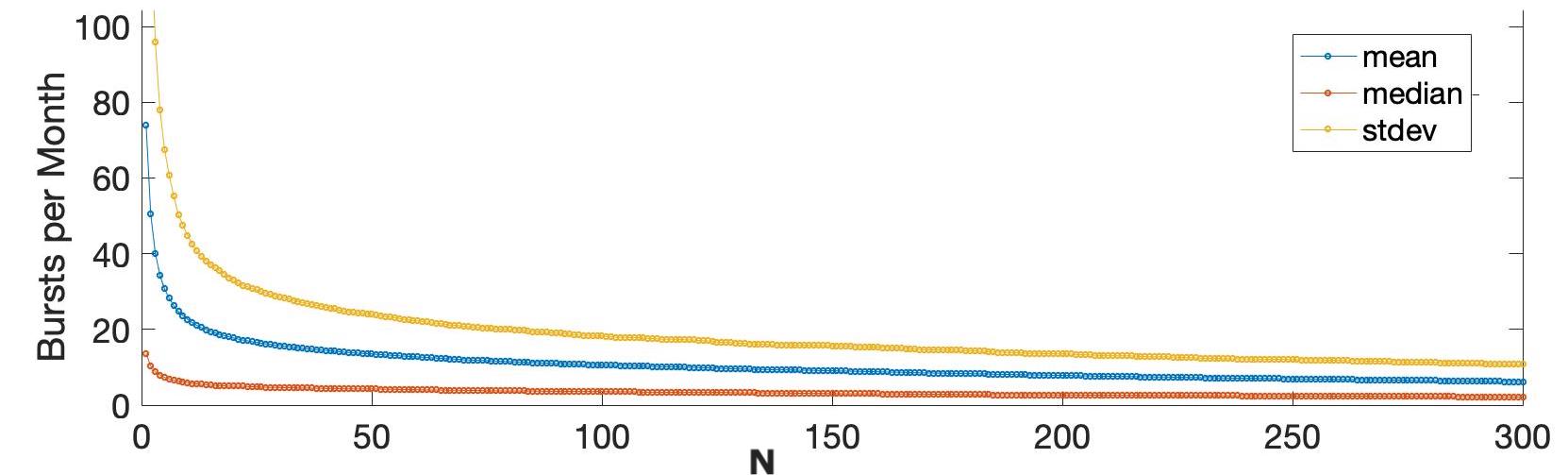}
    \caption{Bursts/month per user as a function of param. $N$.}
    \label{fig:NvsNumBursts}
\end{figure}

Using this definition, we can separate each user's timeline into unique bursts of activity, the key question being how to choose $N$. By measuring the number of bursts per month per user as $N$ varies, we find a steep initial drop with a plateau beginning around $N=40$, with average number of bursts \& standard deviations leveling off around $N=75$ (Fig.~\ref{fig:NvsNumBursts}). Thus, we consider $N \in [75, 100]$. Repeating all ensuing analysis for $N$ at both ends of the range yields similar results indicating robustness of the burst measure with respect to choice to $N$. Furthermore, we find inter-burst times are $4690x$ greater than intra-burst times, indicating that we are not simply confirming the existence of bursts with an arbitrary cutoff of $N$.

\subsubsection{Related work}
Our interest lies in activities that are separated by unusually large intervals (``sustained'' usage), not necessarily those associated with activity rate spikes. Thus, our definition differs from the common interpretation of burst in event streams where focus is on peak activity rate (or spikes)~\cite{al2018wikiburst}. There are also methods for detecting bursts based on an assumed distribution of incoming events~\cite{fung2005parameter} or cost functions~\cite{kleinberg2003bursty}. However, we choose a simple personalized threshold-based method to avoid reliance on additional assumptions while remaining interpretable to mental health forum administrators.
We note that unlike ``sessions'' in web browsing \& search, which are usually bounded by short arbitrary time-intervals (e.g., 30 or 60min~\cite{jansen2007definingsession,halfaker2015usersessions}), bursts are personalized metrics with lengths varying depending on users' post frequency.


\subsection{Quantifying Burstiness of User Activity}
We measure \emph{burstiness} of users based on the fraction of mean intra-burst time (time between a user's own posts within a burst, original or replies), to mean inter-burst time (time between bursts). Then as long as inter-burst time is larger than mean intra-burst post times, burstiness lies between 0 \& 1.
\begin{equation}
     \textrm{Burstiness} = 1 - \left( \frac{\textrm{mean intra-burst time}}{\textrm{mean inter-burst time}}\right) 
\end{equation}

For a user with a consistent activity rate (no bursts), burstiness is defined to be 0. On the other hand, an activity process where burst activity happens at infinitesimal small intra-post times compared to inter-burst gaps, burstiness limits to 1. We can interpret the metric as the difference between observed \& expected ratios ($=1$) of intra-burst and inter-burst times, assuming a nearly uniform activity rate; similar to past work that compared discrepancy of observed activity during a burst to expected activity with no burst~\cite{lappas2009burstiness}. 
Overall, we find that the mean intra-burst time is $2.69$ minutes, compared to average inter-burst time of $9.6$ days. Average bursts on Talklife consist of 20 posts lasting over $6.5$ days.  Across all users, we find an average of $16.6$ bursts and burstiness $\geq 0.99$.



\subsection{Bursts and Mental Health Outcomes}
Assuming people are utilizing Talklife for its intended purpose---a place to seek support in times of mental distress and provide support to others---we can use bursts of activity as a proxy for an individual's progression through an acute problem, or their desire to help others. 
For the remainder of this paper, we show how looking at bursts can provide a fine-grained analysis for a key question in online mental health communities~\cite{Pruksachatkun2019,sharma2018linguistic}: \textbf{Which characteristics of user activity can lead some users to find support and help, and others to fall short?}

To answer this question, we consider two metrics for quantifying positive outcomes: (1) a \textbf{Moment of Cognitive Change} (MOC)~\cite{Pruksachatkun2019} and (2) a user's self-reported \textbf{Change in Mood}.

\subsubsection{Moments of Change}
\input{moc-bursts}

\subsubsection{Change in Mood} \label{sec:mood_change_cat}
Taking mood group 1 (``Sadness'') to be the least positive and group 6 (``Positivity'') to be the most positive, we quantify an individual's change in mood as a progression  where a change of one group in a positive direction yields $= 1$, and a change in a negative direction yields $= -1$. Change in mood is defined as the difference in mood of the last original post in a burst to the first original post. 
While self-reported mood change is an accurate measure of mood across a burst, the metric can be falsely low if an individual experiences an improved mood but proceeds to post only replies to others, since mood is available only for original posts. 


\begin{table}[t]
    \centering \footnotesize
    \caption{Distribution of initial moods in pos. \& neg. bursts}
\label{table:moodgroups_bursts}
\begin{tabular}{p{1.2cm}p{0.6cm}p{0.6cm}p{0.6cm}p{0.6cm}p{0.6cm}p{0.6cm}}
    & \rot[20]{Sadness} & \rot[20]{Inadequacy} & \rot[20]{Frustration} & \rot[20]{Support} & \rot[20]{Relief} & \rot[20]{Positivity} \\
    \hline
    \textbf{Positive} & 57\% & 16\% & 8\% & 6\% & 6\% & 7\% \\
    \hline
    \textbf{Negative} & 55\% & 17\% & 9\% & 6\% & 7\% & 6\% \\
     \hline
\end{tabular}
\end{table}

\subsubsection{Identifying relevant burst characteristics}
In Section~\ref{sec:features}, using \emph{MOC} as the outcome metric, we identify features of bursts which are associated with improved mental health outcomes. Then, in Section~\ref{sec:moodchange}, we evaluate the robustness of our results by computing the correlation of each feature with an independent \textit{Change in Mood} outcome.  As neither of these outcome metrics is perfect, working with two separate metrics allows us to assess robustness of results.


We first identify candidate characteristics using the MOC metric, for the following reasons. First, MOC provides a high-accuracy metric and proxy for cognitive restructuring. While this metric may miss changes in mood which are not explicitly stated, it enables us to identify only the bursts in which a user explicitly states they have benefited \textit{from an interaction}. On the other hand, a user's reported change in mood may be less exact. While we are able to use the first original posts within a burst to identify a general "initial sentiment" for a burst, and determine a "change in mood" across a burst, a user's mood may vary due to experiences off of the platform, resulting in a less-clear boundary for determining a positive or negative post. Additionally, moods are not present for replies, so change can be measures across posts only. 

%% file: moc-bursts.tex
Cognitive restructuring, or cognitive reappraisal, is an emotion regulatory technique worked towards in cognitive-behavioral therapy (CBT)~\cite{gross2003}. This technique involves ``reinterpreting the meaning of a thought or situation to change its emotional trajectory''~\cite{gross2003}, and is particularly powerful as it has had success outside of the clinician's office, including in online mental health support forums~\cite{morris2015,Griffiths2012, OLeary2018}. 

In our work, we utilize a metric developed to identify underlying processes of cognitive restructuring in online mental health communities: the moment of cognitive change (MOC)~\cite{Pruksachatkun2019}. 
A MOC provides a quantitative definition of the psychological concept of an individual expressing a positive change in sentiment for a topic on which they were previously distressed (cognitive restructuring). Since our analysis occurs on the level of bursts, we can track an individual's progression through a series of related posts, threads \& replies, and expand the original definition as such:

\vskip 0.04in
\noindent \textbf{Moment of change:} A positive change in sentiment for a user on a topic that was mentioned by that user in an original post, over the course of a conversation(s) in a single burst. 
\vskip 0.04in

In practice, a MOC is measured  by the occurrence of  phrases in which a user self-reports a positive change.  These phrases include comments such as \textit{"I feel much better now"} and \textit{"I had never thought of that"}, and were validated against manually annotated labels~\cite{Pruksachatkun2019}. These phrases are selected to be high precision, but potentially low recall, in order to avoid false-positives. We use this set of phrases and employ a regular expression-based search to identify MOCs. Note that our measure of MOC may miss positive outcomes from a burst if users do not explicitly state a change in sentiment or express it using language outside of these specific phrases.

%% file: features.tex
Taking MOC as a desirable outcome, we group bursts into positive (contain MOC) and negative (no MOC) subgroups based on whether they contain at least one MOC. Utilizing this grouping, we identify features of bursts that differentiate between positive and negative bursts.
In all cases, the p-value denotes the result of a two-way Kolmogorov-Smirnov test quantifying the distance between the empirical distribution functions of positive and negative subgroups.  

\begin{table}[t]
    \centering \footnotesize
    \caption{Differences in reply-based metrics between MOC positive \& negative bursts. Means and median reported.
    }
\label{Table:engagement}
\begin{tabular}{|p{2.6cm}|p{0.68cm}|p{0.55cm}|p{0.68cm}|p{0.55cm}|p{0.7cm}|}
    \hline
    & \multicolumn{2}{c|}{\textbf{Positive}} & \multicolumn{2}{c|}{\textbf{Negative}} &\\
    \cline{2-5}
      & \textbf{Mean}  & \textbf{Med} & \textbf{Mean}  & \textbf{Med} & \textbf{p}  \\
     \hline
    \textbf{Total replies given} & 7.9 & 5 & 4.3 & 2.6 & 2e-25\\
    \hline
    \textbf{Engagement} & 75\% & 83\% & 50\% & 60\% & 0.05\\
    \hline
    \textbf{Replies received/post} & 4.1 & 3 & 3.8 & 2.8 & 0.02 \\
    \hline
\end{tabular}
\end{table}

\subsection{Meta-analysis}
We find that a MOC occurs on average 68.6\% of the way into a burst, with a mean of 15 posts occurring before a MOC. There is minimal difference between burst length in positive bursts (mean 20.7 posts, median=13) and negative bursts (mean 19 posts, median=12).


Within the data set, 11\% of individuals experience at least one MOC. While coverage across individuals is sizeable, when considering bursts individually, we find MOCs to be more rare, occurring in $1.9\%$ of bursts. This lower rate is consistent with previous analysis on Talklife~\cite{Pruksachatkun2019} and observed difficulty in garnering engagement on the platform: 47\% of posts received 0 replies.



A potential bias could arise if our MOC positive group consists disproportionately of users who post positive, supportive messages. We confirm this is not the case by measuring percentage of posts per mood category in positive and negative MOC groups (Table ~\ref{table:moodgroups_bursts}). Additionally, we note no significant difference in word count between MOC positive (mean $21\pm 18$) and negative ($22\pm25$) bursts indicating "talking more" does not increase likelihood of a MOC. 

\subsection{Engagement with Others}
Online communities such as Talklife create safe spaces for users to seek \& provide support.
Here we analyze the effect of a user's giving \& receiving support, as measured by replies \& engagement.
\begin{equation}
    \textrm{Engagement} = \frac{\textrm{number of replies to others' posts}}{\textrm{total number of own posts}}
    \label{eqn:engagement}
\end{equation}
Based on the number of replies given \& received, the most significant differences between MOC-positive \& negative bursts occur in the extent to which a user engages with others. Those who experience a MOC reply to others at almost twice the rate of those who do not, and thus have higher engagement rates (Table~\ref{Table:engagement}).
Additionally, users who experience a MOC are more likely to start ($51\%$ vs. $34\%$) \& end ($68\%$ vs. $43\%$) a burst on a reply. Notably, the number of replies a user receives has little correlation with MOCs (Table~\ref{Table:engagement}). Therefore, the proportion of time spent engaging with others increases likelihood of an MOC. With respect to beginning on a reply, we hypothesize that users may be searching for a post they relate to \& engaging with others there, prior to posting their own questions. However, an extensive qualitative investigation is needed to confirm this. 

\begin{table}[t]
    \centering \footnotesize
    \caption{Changes in user engagement pre- and post- MOC.}
\label{Table:engagement_prepost}
\begin{tabular}{|p{1.7cm}|p{0.7cm}|p{1cm}|p{0.7cm}|p{1cm}|p{0.7cm}|}
    \hline
      &  \multicolumn{2}{c|}{\textbf{Pre MOC}}  & \multicolumn{2}{c|}{\textbf{Post MOC}} &  \\
     \cline{2-5}
     & \textbf{Mean} & \textbf{Median} & \textbf{Mean} & \textbf{Median} & \textbf{p} \\
     \hline
    \textbf{Engagement} & 71\% & 80\% & 81\% & 91\% & 4e-10 \\
    \hline
\end{tabular}
\end{table}

Taking MOC as a marker of behavior change, we can measure a user's engagement pre- and post- MOC. We find that users increase engagement almost 10\% after receiving an MOC, potentially indicating that users who experience a MOC are more likely to want to give back to the community by being a peer supporter (Table~\ref{Table:engagement_prepost}). 

\begin{figure*}[t]
    \centering
    \includegraphics[width=0.45\textwidth]{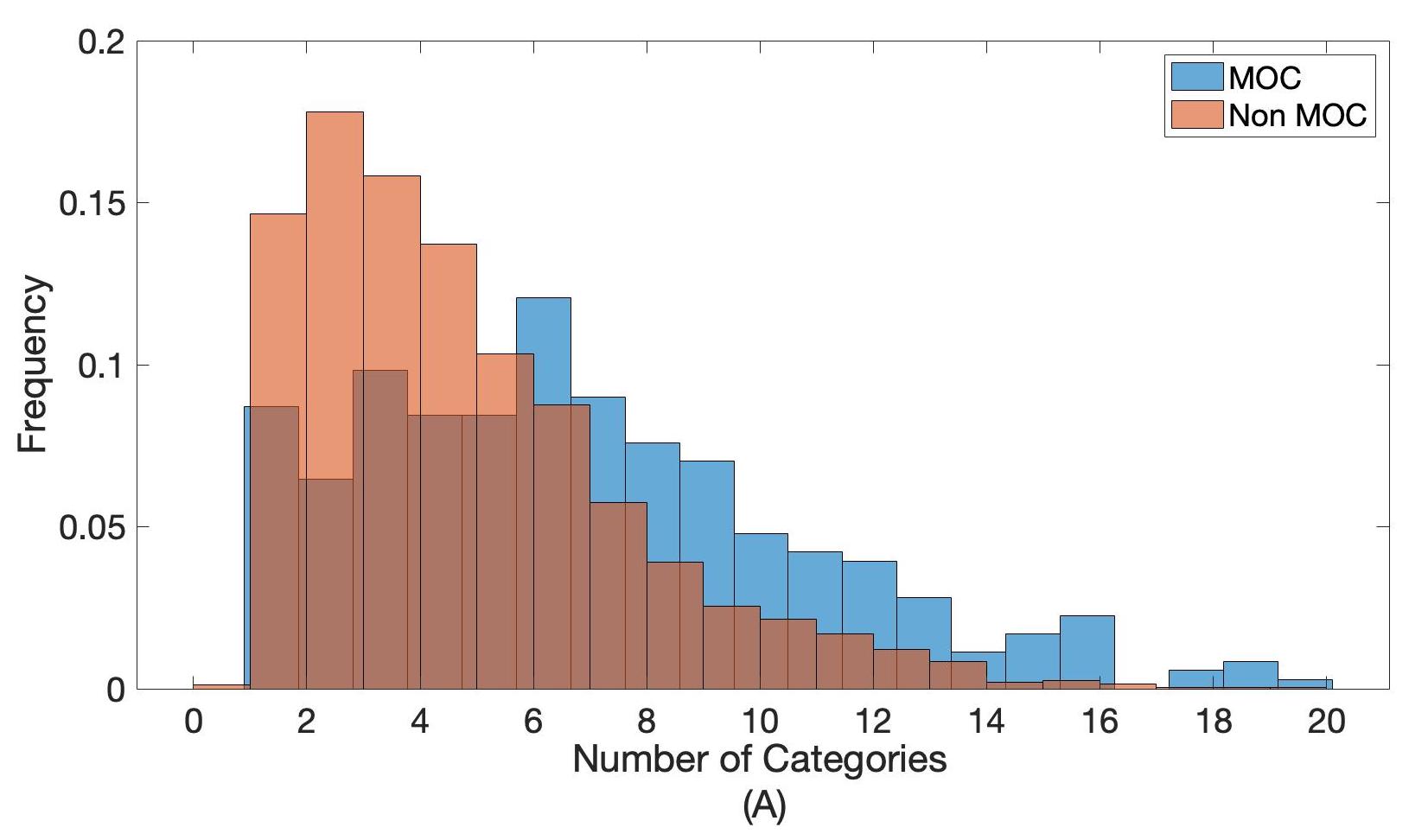}     \includegraphics[width=0.45\textwidth]{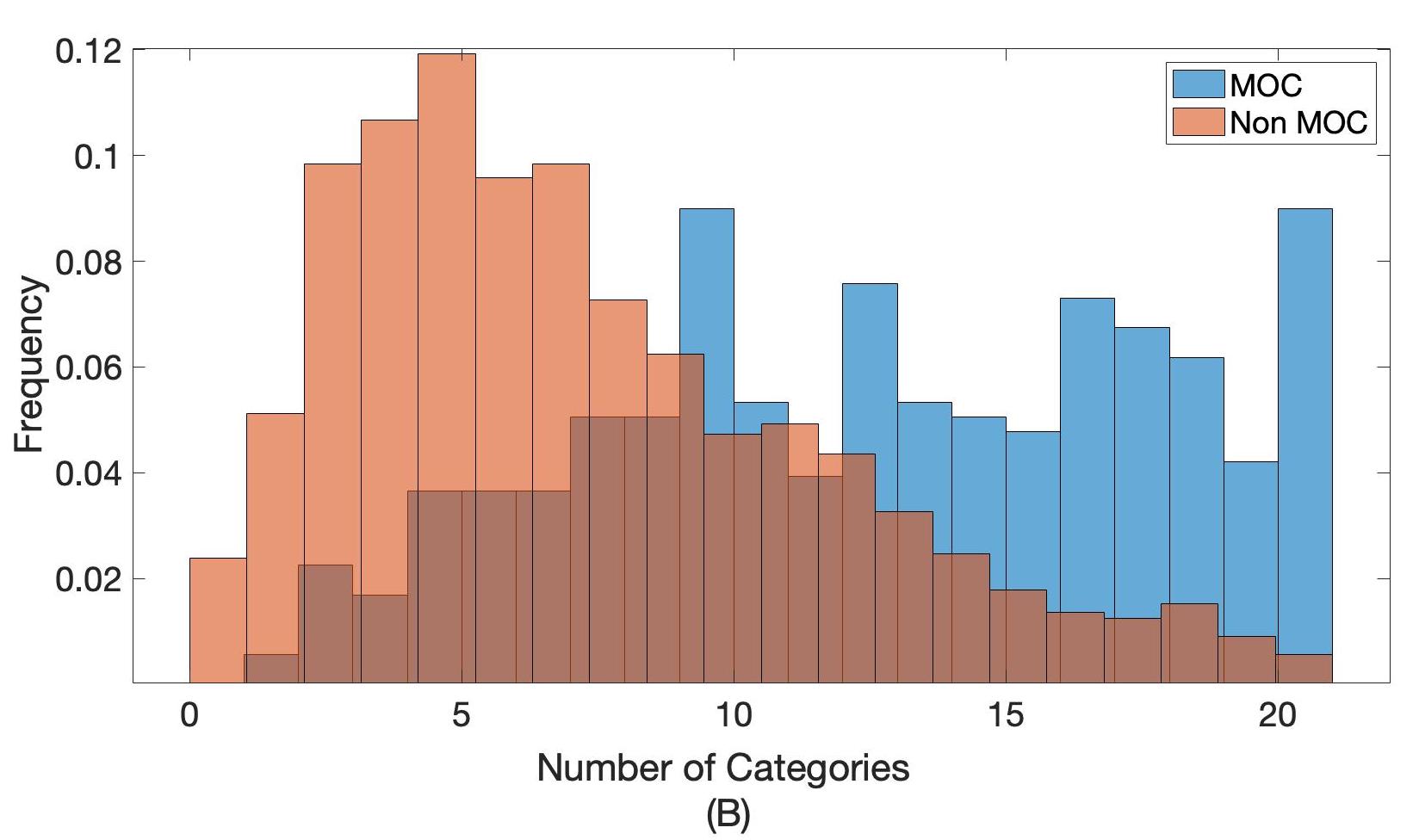}
    \caption{Average number of categories users post in for positive and negative MOC bursts: (A) Posts, (B) Replies.}
    \label{fig:hist_cat}
\end{figure*}


\subsection{Diversity in Post Categories}

Categories provide one of the only filters for users when viewing their timelines on Talklife, as users can select to view only posts from selected categories. While we cannot measure if users are filtering categories directly, we can measure the breadth of categories in which they post. We can then quantify benefit of higher or lower breadth \& similarity of categories between posts \& replies.
Users who experience MOCs are more likely to have increased diversity in categories for both original posts (mean 6.7 vs. 4.3, $p=2\times 10^{-26}$) \& replies (12.3 .vs 7.4, $p=4.5\times 10^{-47}$). When considering distributions, we find that the greater shift occurs in user replies (Fig.~\ref{fig:hist_cat}). 

However, it is not simply about posting in multiple categories, but rather posting \& replying in similar categories. Here we consider the effect of \emph{post similarity}, taken to be the cosine distance between overall category proportions of replies \& posts. We utilize this measure as a proxy for users looking for a smaller community within Talklife by identifying a specific category which is relevant to them, then reading questions \& replying to posts within said category. 
We find users who consistently post a majority of posts \& replies to similar categories increase their likelihood of experiencing a MOC in a burst, with probability of MOC being $0.37$ when top categories match, versus $0.17$ otherwise. These probabilities hold after matching users on number of post \& reply categories.

\subsection{Type of Support Received}
Building on the work of Andalibi et al.~\cite{Andalibi2017} on support types in social media, we now examine the effect of providing various modes of support to others. We identify two categories most relevant to Talklife: emotional support, and network and instrumental support. 

To determine if complexity of responses matter, we break down emotional support into (1) Simple Emotional Support (SES) \& (2) Complex Emotional Support (CES).
For SES, we consider posts matching exact simple phrases identified by Andalibi et al. such as \textit{"Same", "Been there"} and \textit{"I know how you feel"}. CES includes the same SES phrases, but with added context. For example, \textit{"I know how you feel. Last year, I was also ..."}. We find users who experience a MOC within a burst are more likely to both give \& receive CES, but there is little difference in SES given or received (Table.~\ref{Table:MOCfeatures}). This suggests users benefit more from giving \& receiving longer, more thoughtful replies rather than simple acknowledgements of support. The strongest difference between positive \& negative groups is in CES a user \emph{gives}, suggesting that by taking time to engage in thoughtful supportive interactions with others, a user's own well-being can benefit. Still, it could be that a user gives CES support after having a MOC in a burst, rather than before. To refute this hypothesis, we divide user activity pre- \& post-experiencing a MOC \& evaluate the difference between support given pre \& post MOC. We find no significant difference ($p =1$ for SES, \& $0.95$ for CES). 

\begin{table}[t]
    \centering \footnotesize
    \caption{Mean support type in MOC pos. \& neg. bursts.}
\label{Table:MOCfeatures}
\begin{tabular}{p{1.2cm}p{1cm}p{1cm}p{1cm}p{1cm}}
     & \rot[14]{\textbf{SES Received}} & \rot[14]{\textbf{CES Received}} & \rot[14]{\textbf{SES Given}} & \rot[14]{\textbf{CES Given}}\\
     \hline
     \textbf{Positive} & 0.0013 & 0.029 & 0.0009 & 3.3e-6  \\
     \hline
     \textbf{Negative} & 0.002 & 0.018 & 0.0016 & 1.5e-7 \\
     \hline
     {p-value} & 1 & 0.60 & 1 & 0.04  \\
    \hline
\end{tabular}
\end{table}

Next we consider the second support category (3) Network \& Instrumental support (NS), which we can use as a proxy for support for which we do not have data: that which may occur within direct messages on TalkLife or off the platform. Phrases utilized include \textit{"DM me"} \& \textit{"Message me on kik"}.
We find no significant difference in NS given or received between positive and negative groups, ($p=1$ for both). Furthermore, we find NS is present in fewer than 0.3\% of all posts, indicating that any off-platform MOC events would be unlikely to affect our overall results.



\subsection{Linguistic Features}
To further analyze the types of language used, we employ the Linguistic Inquiry \& Word Count (LIWC) tool (2015 version) \cite{Tausczik_2009}. LIWC is a psycholinguistic text analysis tool which draws categories based on expressive writing literature that associates language with therapeutic symptoms. It has been validated for predictive tasks in a variety of mental health contexts ~\cite{dechoudhury2013, Saha_2017}.
Using LIWC, we determine the proportion of affective (positive \& negative affect) keywords utilized by users throughout a burst. Values reported are mean LIWC category scores.

Overall, users who experience a MOC are more likely to: use positive affective language ($1.3$ vs. $0.36$, $p=9\times 10^{-29}$), receive replies from others containing positive affective words ($0.49$ vs. $0.2$, $p=2.2\times 10^{-13}$), and use fewer negative affective words (3.3e-06 vs. 1.5e-07, $p=0.04$). 
With respect to negative affect in replies \emph{from} others, we find no significant differences between positive and negative bursts (0.0013 vs 0.002, $p=1$). Additionally, when we compare language composition of posts pre- and post- experiencing a MOC, there is no significant changes in positive or negative affect.






%% file: user_benefit.tex
Taking features identified as significantly higher in MOC bursts, we check robustness by removing the MOC stipulation, and instead conditioning bursts on a found feature to determine if a secondary outcome (mood change,~\ref{sec:mood_change_cat}) may be achieved regardless of MOC.

\subsection{Robust Actionable Features}

\subsubsection{Post Persistence} 
On average, MOCs occurs 15 posts into a burst. Disregarding MOC, if a user posts $\geq15$ times within a burst, we find a 13-fold increase in mood change. Taking 15+ posts to be "long", we find a mean $0.13$ increase in mood in long bursts, vs. a $0.01$ increase in short bursts.
Furthermore, the likelihood that a mood change is \emph{positive} (non-zero) increases 3.4-fold ($0.07$ vs. $0.23$). 
\subsubsection{Giving Complex Emotional Support}
Regardless of a MOC, we find individuals who support others with CES experience an average 5-fold increase in mood change ($0.14$ vs. $0.023$). Interestingly, although we found SES to not differ in MOC positive \& negative subgroups, we do find that individuals who offer simple emotional support have a 3.5x increased average mood change ($0.11$ vs. $0.03$).

\subsection{Non-Robust Features}
Although increased engagement and positive affective language varied significantly between positive and  negative MOC groups, neither feature was robust to the mood-change outcome. While mood increased slightly for users with high engagement (rates above the MOC-correlated value of 75\%) than mean engagement (51\%), and for those using positive affective language above mean levels, overall mood change remained near zero ($<0.08$ in all cases). 

%% file: implications_discussion.tex
To our knowledge, this is the first work to characterize medium-term activity in online mental health communities. It provides a novel framework for analysis of user behavior, gives a proxy for user's progressions through acute problems, \& bridges the gap between prior work on short~\cite{Pruksachatkun2019} \& long-term scales~\cite{park2017longitudinal, de2016discovering}. 


The identified features of persistence and giving complex emotional support have potential for use by mental health platform owners to improve user experience. For example, suggesting users to provide complex emotional support rather than simple support. However, we caution that these results report correlations,
not causal factors.
In future work we aim to further this analysis by establishing \emph{causal} relationships between burst features \& positive mental health changes, as well as studying changes happening \emph{between} bursts. 

Finally, we believe the implications of such behavior patterns can extend to how we think about churn \& user retention. Unlike general social media platforms, it is not necessary that a user exiting from a mental health platform is dissatisfied. On the contrary, our results show that users may be turning to a platform such as Talklife at times of need, and taking breaks at other times. While we cannot definitively say when or why a user may return, formulating the problem using bursts \& breaks can help better track churn. 

\vskip 0.04in
\noindent \textbf{Acknowledgements:} We thank Monojit Choudhury, Koustuv Saha \& Sachin Pendse for their valuable feedback on the work and Talklife for providing access to the data.

%% file: main.bbl
\begin{thebibliography}{10}

\bibitem{al2018wikiburst}
Reham Al~Tamime, Richard Giordano, and Wendy Hall.
\newblock Observing burstiness in wikipedia articles during new disease
  outbreaks.
\newblock In {\em Proceedings of the 10th ACM Conference on Web Science}, pages
  117--126. ACM, 2018.

\bibitem{Andalibi2017}
Nazanin Andalibi, Pinar Ozturk, and Andrea Forte.
\newblock Sensitive self-disclosures, responses, and social support on
  instagram.
\newblock {\em Proceedings of the 2017 ACM Conference on Computer Supported
  Cooperative Work and Social Computing - CSCW '17}, 2017.

\bibitem{dechoudhury2013}
Munmum~De Choudhury, Michael Gamon, Scott Counts, and Eric Horovitz.
\newblock Predicting depression via social media.
\newblock {\em AAAI Conference on Weblogs and Social Media}, 2013.

\bibitem{dasgupta2008social}
Koustuv Dasgupta, Rahul Singh, Balaji Viswanathan, Dipanjan Chakraborty,
  Sougata Mukherjea, Amit~A Nanavati, and Anupam Joshi.
\newblock Social ties and their relevance to churn in mobile telecom networks.
\newblock In {\em Proceedings of the 11th international conference on Extending
  database technology: Advances in database technology}, pages 668--677. ACM,
  2008.

\bibitem{de2014mentaldisclosure}
Munmun De~Choudhury and Sushovan De.
\newblock Mental health discourse on reddit: Self-disclosure, social support,
  and anonymity.
\newblock In {\em Eighth International AAAI Conference on Weblogs and Social
  Media}, 2014.

\bibitem{de2013predicting}
Munmun De~Choudhury, Michael Gamon, Scott Counts, and Eric Horvitz.
\newblock Predicting depression via social media.
\newblock In {\em Seventh international AAAI conference on weblogs and social
  media}, 2013.

\bibitem{de2017language}
Munmun De~Choudhury and Emre Kiciman.
\newblock The language of social support in social media and its effect on
  suicidal ideation risk.
\newblock In {\em Eleventh International AAAI Conference on Web and Social
  Media}, 2017.

\bibitem{de2016discovering}
Munmun De~Choudhury, Emre Kiciman, Mark Dredze, Glen Coppersmith, and Mrinal
  Kumar.
\newblock Discovering shifts to suicidal ideation from mental health content in
  social media.
\newblock In {\em Proceedings of the 2016 CHI conference on human factors in
  computing systems}, pages 2098--2110, 2016.

\bibitem{dror2012churnyahoo}
Gideon Dror, Dan Pelleg, Oleg Rokhlenko, and Idan Szpektor.
\newblock Churn prediction in new users of yahoo! answers.
\newblock In {\em Proceedings of the 21st International Conference on World
  Wide Web}, pages 829--834. ACM, 2012.

\bibitem{ernala2017linguistic}
Sindhu~Kiranmai Ernala, Asra~F Rizvi, Michael~L Birnbaum, John~M Kane, and
  Munmun De~Choudhury.
\newblock Linguistic markers indicating therapeutic outcomes of social media
  disclosures of schizophrenia.
\newblock {\em Proceedings of the ACM on Human-Computer Interaction},
  1(CSCW):43, 2017.

\bibitem{fung2005parameter}
Gabriel Pui~Cheong Fung, Jeffrey~Xu Yu, Philip~S Yu, and Hongjun Lu.
\newblock Parameter free bursty events detection in text streams.
\newblock In {\em Proceedings of the 31st international conference on Very
  large data bases}, pages 181--192. VLDB Endowment, 2005.

\bibitem{golder2007rhythms}
Scott~A Golder, Dennis~M Wilkinson, and Bernardo~A Huberman.
\newblock Rhythms of social interaction: Messaging within a massive online
  network.
\newblock In {\em Communities and technologies 2007}, pages 41--66. Springer,
  2007.

\bibitem{grafarend2012}
Erik Grafarend and Joseph Awange.
\newblock {\em Linear and Nonlinear Models: Fixed effects, random effects, and
  total least squares}.
\newblock Springer, 2012.

\bibitem{Griffiths2012}
Kathleen~M. Griffiths, Andrew~J. Mackinnon, Dimity~A. Crisp, Helen Christensen,
  Kylie Bennett, and Louise Farrer.
\newblock The effectiveness of an online support group for members of the
  community with depression: A randomised controlled trial.
\newblock {\em PLoS ONE}, 7(12):e53244, Dec 2012.

\bibitem{gross2003}
J.~J. Gross and O.~P. John.
\newblock Individual differences in two emotion regulation processes:
  Implications for affect, relationships, and well-being.
\newblock {\em Journal of Personality and Social Psychology}, 85(2):348--462,
  2003.

\bibitem{guille2012predictive}
Adrien Guille and Hakim Hacid.
\newblock A predictive model for the temporal dynamics of information diffusion
  in online social networks.
\newblock In {\em Proceedings of the 21st international conference on World
  Wide Web}, pages 1145--1152. ACM, 2012.

\bibitem{guo2009userpatterns}
Lei Guo, Enhua Tan, Songqing Chen, Xiaodong Zhang, and Yihong~Eric Zhao.
\newblock Analyzing patterns of user content generation in online social
  networks.
\newblock In {\em Proceedings of the 15th ACM SIGKDD international conference
  on Knowledge discovery and data mining}, pages 369--378, 2009.

\bibitem{halfaker2015usersessions}
Aaron Halfaker, Os~Keyes, Daniel Kluver, Jacob Thebault-Spieker, Tien Nguyen,
  Kenneth Shores, Anuradha Uduwage, and Morten Warncke-Wang.
\newblock User session identification based on strong regularities in
  inter-activity time.
\newblock In {\em Proceedings of the 24th International Conference on World
  Wide Web}, pages 410--418. International World Wide Web Conferences Steering
  Committee, 2015.

\bibitem{jamali2011modeling}
Mohsen Jamali, Gholamreza Haffari, and Martin Ester.
\newblock Modeling the temporal dynamics of social rating networks using
  bidirectional effects of social relations and rating patterns.
\newblock In {\em Proceedings of the 20th international conference on World
  wide web}, pages 527--536. ACM, 2011.

\bibitem{jansen2007definingsession}
Bernard~J Jansen, Amanda Spink, Chris Blakely, and Sherry Koshman.
\newblock Defining a session on web search engines.
\newblock {\em Journal of the American Society for Information Science and
  Technology}, 58(6):862--871, 2007.

\bibitem{kleinberg2003bursty}
Jon Kleinberg.
\newblock Bursty and hierarchical structure in streams.
\newblock {\em Data Mining and Knowledge Discovery}, 7(4):373--397, 2003.

\bibitem{kumar2005bursty}
Ravi Kumar, Jasmine Novak, Prabhakar Raghavan, and Andrew Tomkins.
\newblock On the bursty evolution of blogspace.
\newblock {\em World Wide Web}, 8(2):159--178, 2005.

\bibitem{kumar2010structure}
Ravi Kumar, Jasmine Novak, and Andrew Tomkins.
\newblock Structure and evolution of online social networks.
\newblock In {\em Link mining: models, algorithms, and applications}, pages
  337--357. Springer, 2010.

\bibitem{lappas2009burstiness}
Theodoros Lappas, Benjamin Arai, Manolis Platakis, Dimitrios Kotsakos, and
  Dimitrios Gunopulos.
\newblock On burstiness-aware search for document sequences.
\newblock In {\em Proceedings of the 15th ACM SIGKDD international conference
  on Knowledge discovery and data mining}, pages 477--486. ACM, 2009.

\bibitem{lancetdepression}
Gin S et~al. Malhi.
\newblock {\em The Lancet}, 392, 2018.

\bibitem{morris2015}
Robert~R Morris, Stephen~M Schueller, and Rosalind~W Picard.
\newblock Efficacy of a web-based, crowdsourced peer-to-peer cognitive
  reappraisal platform for depression: Randomized controlled trial.
\newblock {\em J Med Internet Res}, 17(3):e72, Mar 2015.

\bibitem{naslund2016future}
JA~Naslund, KA~Aschbrenner, LA~Marsch, and SJ~Bartels.
\newblock The future of mental health care: peer-to-peer support and social
  media.
\newblock {\em Epidemiology and psychiatric sciences}, 25(2):113--122, 2016.

\bibitem{OLeary2018}
Kathleen O'Leary, Stephen~M. Schueller, Jacob~O. Wobbrock, and Wanda Pratt.
\newblock ``suddenly, we got to become therapists for each other''.
\newblock {\em Proceedings of the 2018 CHI Conference on Human Factors in
  Computing Systems - CHI '18}, 2018.

\bibitem{pal2012evolutionexperts}
Aditya Pal, Shuo Chang, and Joseph~A Konstan.
\newblock Evolution of experts in question answering communities.
\newblock In {\em sixth international AAAI conference on weblogs and social
  media}, 2012.

\bibitem{park2017longitudinal}
Albert Park and Mike Conway.
\newblock Longitudinal changes in psychological states in online health
  community members: Understanding the long-term effects of participating in an
  online depression community.
\newblock {\em J Med Internet Res}, 19(3):e71, Mar 2017.

\bibitem{patel2018lancet}
Vikram Patel, Shekhar Saxena, Crick Lund, Graham Thornicroft, Florence
  Baingana, Paul Bolton, Dan Chisholm, Pamela~Y Collins, Janice~L Cooper,
  Julian Eaton, et~al.
\newblock The lancet commission on global mental health and sustainable
  development.
\newblock {\em The Lancet}, 392(10157):1553--1598, 2018.

\bibitem{Pruksachatkun2019}
Yada Pruksachatkun, Sachin~R. Pendse, and Amit Sharma.
\newblock Moments of change.
\newblock {\em Proceedings of the 2019 CHI Conference on Human Factors in
  Computing Systems - CHI '19}, 2019.

\bibitem{pudipeddi2014user}
Jagat~Sastry Pudipeddi, Leman Akoglu, and Hanghang Tong.
\newblock User churn in focused question answering sites: characterizations and
  prediction.
\newblock In {\em Proceedings of the 23rd International Conference on World
  Wide Web}, pages 469--474. ACM, 2014.

\bibitem{Saha_2017}
Koustuv Saha and Munmun De~Choudhury.
\newblock Modeling stress with social media around incidents of gun violence on
  college campuses.
\newblock {\em Proceedings of the ACM on Human-Computer Interaction},
  1(CSCW):1--27, Dec 2017.

\bibitem{sharma2018linguistic}
Eva Sharma and Munmun De~Choudhury.
\newblock Mental health support and its relationship to linguistic
  accommodation in online communities.
\newblock In {\em Proceedings of the 2018 CHI Conference on Human Factors in
  Computing Systems}, page 641. ACM, 2018.

\bibitem{talklife}
Talklife.
\newblock https://talklife.co/.

\bibitem{Tausczik_2009}
Yla~R. Tausczik and James~W. Pennebaker.
\newblock The psychological meaning of words: Liwc and computerized text
  analysis methods.
\newblock {\em Journal of Language and Social Psychology}, 29(1):24--54, Dec
  2009.

\bibitem{tsugawa2015recognizing}
Sho Tsugawa, Yusuke Kikuchi, Fumio Kishino, Kosuke Nakajima, Yuichi Itoh, and
  Hiroyuki Ohsaki.
\newblock Recognizing depression from twitter activity.
\newblock In {\em Proceedings of the 33rd annual ACM conference on human
  factors in computing systems}, pages 3187--3196. ACM, 2015.

\bibitem{wei2010reviewrfm}
Jo-Ting Wei, Shih-Yen Lin, and Hsin-Hung Wu.
\newblock A review of the application of rfm model.
\newblock {\em African Journal of Business Management}, 4(19):4199--4206, 2010.

\bibitem{white2009cyberchondria}
Ryen~W White and Eric Horvitz.
\newblock Cyberchondria: studies of the escalation of medical concerns in web
  search.
\newblock {\em ACM Transactions on Information Systems (TOIS)}, 27(4):23, 2009.

\bibitem{yang2011temporalpatterns}
Jaewon Yang and Jure Leskovec.
\newblock Patterns of temporal variation in online media.
\newblock In {\em Proceedings of the fourth ACM international conference on Web
  search and data mining}, pages 177--186. ACM, 2011.

\bibitem{zhao2012eventbursts}
Wayne~Xin Zhao, Baihan Shu, Jing Jiang, Yang Song, Hongfei Yan, and Xiaoming
  Li.
\newblock Identifying event-related bursts via social media activities.
\newblock In {\em Proceedings of the 2012 Joint Conference on Empirical Methods
  in Natural Language Processing and Computational Natural Language Learning},
  pages 1466--1477. Association for Computational Linguistics, 2012.

\end{thebibliography}
